# Macroscopic, layered onion shell like magnetic domain structure generated in YIG film using ultrashort, megagauss magnetic pulses

Kamalika Nath[1], P. C. Mahato[1], Moniruzzaman Shaikh[2], Kamalesh Jana[2], Amit D Lad[2], Deep Sarkar[2], Rajdeep Sensarma[3], G. Ravindra Kumar[2#], S. S. Banerjee[1*]

[1]Department of Physics, Indian Institute of Technology, Kanpur, Uttar Pradesh, India,
[2]Department of Nuclear and Atomic Physics, Tata Institute of Fundamental Research, Homi Bhabha Road, Mumbai, India.
[3]Department of Theoretical Physics, Tata Institute of Fundamental Research, Homi Bhabha Road, Mumbai, India.

Corresponding authors' email: [#]grk@tifr.res.in, [*]satyajit@iitk.ac.in.

**Abstract:** Study of the formation and evolution of large scale, ordered structures is an enduring theme in science. The generation, evolution and control of large sized magnetic domains are intriguing and challenging tasks, given the complex nature of competing interactions present in any magnetic system. Here, we demonstrate large scale non-coplanar ordering of spins, driven by picosecond, megagauss magnetic pulses derived from a high intensity, femtosecond laser. Our studies on a specially designed Yttrium Iron Garnet (YIG)/dielectric/metal film sandwich target, show the creation of complex, large, concentric, elliptical shaped magnetic domains which resemble the layered shell structure of an onion. The largest shell has a major axis of over hundreds of micrometers, in stark contrast to conventional sub micrometer scale polygonal, striped or bubble shaped magnetic domains found in magnetic materials, or the large dumbbell shaped domains produced in magnetic films irradiated with accelerator based relativistic electron beams. Through micromagnetic simulations, we show that the giant magnetic field pulses create ultrafast terahertz (THz) spin waves. A snapshot of these fast propagating spin waves is **stored as the layered onion shell shaped domains in the YIG film. Typically, information transport via spin waves in magnonic devices occurs in the gigahertz (GHz) regime, where the devices are susceptible to thermal disturbances at room temperature. Our intense laser light pulse - YIG sandwich target combination, paves the way for room temperature table-top THz spin wave devices, which operate just above or in the range of the thermal noise floor. This dissipation-less device offers ultrafast control of spin ~~in~~formation over distances of few hundreds of microns.**

Conventional condensed matter systems display a diverse variety of static magnetization configurations like Bloch or Neel domain walls, magnetic vortices (1,2), stripe domains (3) and skrymions (4). These domains arise out of a competition between different magnetic exchange and magnetostatic energies (5,6). Over the past few decades, a great deal of effort has gone into using light for probing different aspects of condensed matter physics (7,8,9,10,11,12). Effects of light interacting with magnetism have been primarily studied with low intensity ($I \sim 10^5$-$10^6$ Wcm$^{-2}$) femtosecond (fs) lasers (7,8,9,10) for exploring demagnetization processes occurring on time scales of a few picoseconds (ps) (10,13,14,15). Recently, optical coupling of angular momentum of light with spins in magnetizable media has been shown to create micron-sized domains on fs timescales (8,16,17,18). The use of high intensity femtosecond laser pulses ($I \sim 10^{14}$ - $10^{18}$ Wcm$^{-2}$) for such studies has however not been attempted so far. This may be attributed to the apprehension that the enormous energy scale associated with such excitation would overwhelm the spin - spin interaction energy scale and the thermal damage induced by such intense laser pulse would obliterate the possibility of seeing any ordered spin configuration. Indeed, direct irradiation with such intense pulses typically ablates the material creating a high temperature plasma. However, interaction of an intense fs laser pulse with the plasma is interesting as it is known to produce giant megagauss (MG) magnetic field pulses of picosecond duration (19,20,21,22,23). It is therefore worthwhile studying the response of magnetic materials to such intense magnetic pulses. In this paper, with an innovative target design and careful control of experimental conditions, we demonstrate the creation of novel, unusual spin structures created by this magnetic pulse.

Here we study the response of Yttrium Iron Garnet (YIG) film subjected to megagauss magnetic field pulses produced by the interaction of a few hundred petawatt/cm$^2$ intensity, 30 fs laser pulse with a solid target. YIG is a well-known ferrimagnetic insulator film with very low damping, which in recent times has become an attractive material for studying magnon dynamics (24,25,26) (magnons are quasi particles associated with spin waves). Low damping of magnetization dynamics coupled with large magnon diffusion lengths reaching several microns (27), make YIG an important material for applications in magnonics (28,29,30), spin calorimetrics (31,32,33) and magnon-based microwave applications (34,35). A careful target design is however, crucial for eliminating the ablative degradation of YIG due to laser induced ionization and subsequent heating. We therefore implement a novel sandwich target geometry of metal film (Al)-dielectric-YIG (Fig. 1*A*), where the laser irradiates the top Al layer, leaving the lower YIG layer unaffected by the laser induced damage. Magneto-optical microscopy (MOM) of the YIG samples exposed to the laser generated giant magnetic field shows the creation of novel, large, concentric, elliptically shaped magnetic domains extending up to a few hundreds of microns from the projected irradiation location. The shape resembles layered shells of an onion. Furthermore, we see that the local magnetic field direction flips up and down periodically across these elliptical domain structures and its magnitude also has a periodic variation with distance from the center of the irradiation. Micromagnetic simulation of a YIG film subjected to megagauss field pulse shows the excitation of ripples of spin waves travelling across the low damping YIG film, a few picoseconds after the pulse. The spin waves cause moments to gradually rotate out of the film plane periodically resulting in the observed behavior of the measured local field. These fast spin waves diffuse up to a few hundreds of microns in the YIG film from the projected laser irradiation site, giving rise to a non-collinear spin configuration, which in turn we propose, gives rise to an additional Dzyloshinskii–Moriya type interaction contribution to the magnetic energy of YIG. This interaction together with pinning effects, results in the spin waves getting stored as the layered onion shell like magnetic domain structure in YIG.

Each target (Fig. 1*A*) consists of a 16 μm thick Al film suspended over a GGG (Gallium Gadolinium Garnet) substrate in a sandwich configuration. The lower side of the GGG substrate has a Bismuth doped YIG film grown on it (36,37). We use single pulses of 25 femtosecond (fs) laser (p-polarized, center wavelength 800 nm) having 20 μm beam diameter to irradiate identical points at different locations on the Al layer at an angle of incidence of 45°. The laser intensities used are between $3 \times 10^{17}$ to $1 \times 10^{18}$ Wcm$^{-2}$ (details of setup in Supplementary section). The YIG film is isolated from the optical field of the intense laser as well the heating effects it generates. The sandwich configuration provides a two level protection to the YIG film. Firstly, it eliminates laser induced ionization of the YIG film and the resulting thermal heat load that could lead to direct damage of the film, since the intense laser pulse ablates the sacrificial Al layer which takes away these deleterious effects. Secondly, the magnetic YIG layer has additional shielding from the heating effects provided by the intervening 200 μm thick dielectric air gap and the GGG layer present between Al and YIG film layer. The YIG film was devoid of any micron sized magnetic domains prior to irradiation. As late as four days after irradiation, the irradiated region is imaged using a high sensitivity magneto-optic microscope (MOM setup details in Supplementary information and Ref. 23). The magneto-optical intensity is $\propto B_z^2$, where $B_z$ is the component of local field perpendicular to the surface. The $B_z(x,y)$ distribution is determined from the MOM intensity distribution by suitable calibration (23) (the *x* and *y* axes are in the film plane while *z* axis is perpendicular to the film).

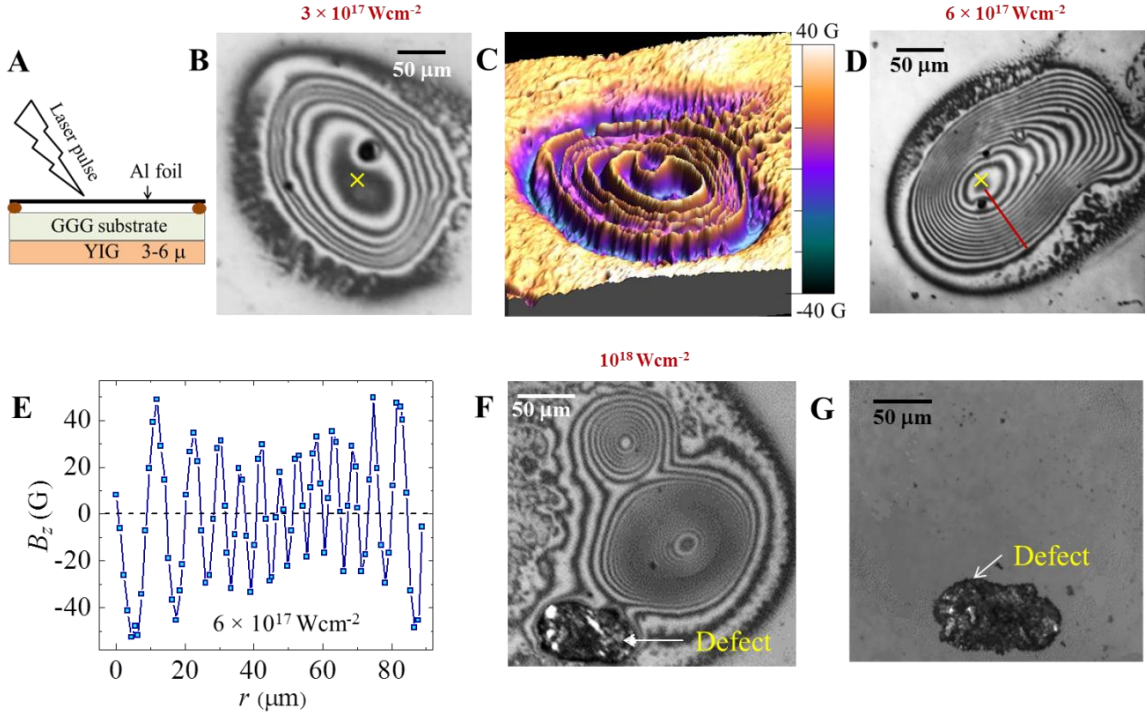

**Fig. 1,** MOM Images of laser Irradiated YIG films. (*A*), Schematic shows the incidence of the femtosecond laser pulse on a 16 μm thick Al film suspended 100 μm above the GGG substrate, on the lower side of which is the YIG film. A dielectric layer of thickness ~ 200 μm composed of air and the GGG substrate, exists between the Al and the YIG film. The × mark is the estimated projected location of the laser spot on the YIG film. (*B,D,F*) MOM images of concentric elliptical magnetic domains generated in YIG film are shown after irradiation at laser intensities (I) of $3 \times 10^{17}$ Wcm$^{-2}$, $6 \times 10^{17}$ Wcm$^{-2}$ and $10^{18}$ Wcm$^{-2}$, respectively. (*C*) Color coded three-dimensional map of the $B_z(x,y)$ distribution in the elliptical domains of 1(*B*). The colors represent the magnitude of the $B_z$ values as shown in the color-bar scale. (*E*) A one-dimensional map of $B_z$ profile (viz., $B_z(r)$) measured along the red line in Fig. 1(*D*), shows periodic oscillations in $B_z$ as one traverses the concentric rings of the elliptical domain pattern. The maximum amplitude of the oscillating $B_z(r)$ is ± 40 G. (*F*) Shows two domain patterns generated in YIG with laser pulse irradiation of intensity of $10^{18}$ Wcm$^{-2}$. The smaller upper domain structure is more circular than the one below. Also seen is a defect in the YIG film which was present before irradiation. (*G*) MOM image of the region same as that in (*F*), imaged after 10 days of laser irradiation, with the black defect as the identifier. Here we see layered onion shell like magnetic domain patterns have disappeared.

The MOM images in Figs. 1 *B*, *D* and *F* show the creation of magnetic domain patterns recorded in the YIG film layer, after irradiation by single pulses of laser intensities of $3 \times 10^{17}$ Wcm$^{-2}$, $6 \times 10^{17}$ Wcm$^{-2}$ and $10^{18}$ Wcm$^{-2}$, respectively. All images show the formation of concentric black and white elliptical shaped onion shell-like magnetic domains in the YIG film. The overall sizes of the elliptic domain patterns at different laser intensities are comparable and do not scale with laser intensity. The number of concentric rings however increases with increasing laser intensity. The full extent of the concentric domain patterns is at least an order of magnitude larger than the laser beam diameter of 20 μm. The peaks and troughs in the color-coded three-dimensional view of $B_z(x,y)$ in Fig. 1C, correspond to periodic modulation of local field $B_z$ across the bright and dark regions of the pattern in Fig. 1B. A comparison of the concentric domains in Fig. 1B with Fig. 1D shows the number of concentric rings increasing with intensity (*I*) of the laser. Furthermore, Fig. 1F shows that at a high intensity of $10^{18}$ Wcm$^{-2}$ there is a large elliptical domain pattern with concentric rings, and an adjacent satellite concentric domain with a relatively lower eccentricity. Fig. 1G shows a MOM image of the same region of the YIG film as in Fig. 1F recorded 10 days after irradiation. Note that image of the defect in the YIG film layer of Fig. 1F is retained in Fig. 1G, however the magneto-optical contrast of the domain patterns has diminished such that the patterns are no longer discernible. This disappearance of the pattern is naturally expected, as the remnant magnetization of the film at room temperature decays out to zero with time. This feature suggests the domain patterns are not permanent and irreversible, i.e., they are not related to laser induced heating damage. The magnetized regions are generated by the action of the megagauss magnetic pulse created by the laser. The following sections explore the origin of this quasi-stable domain feature observed in YIG film.

It is well established in intense laser-solid interaction studies that a high intensity, p-polarized femtosecond laser pulse incident on a target at a non-normal angle sets up electron waves in the generated plasma, which grow to large amplitude before breaking. This breaking unleashes a giant current pulse (~mega-ampere) that travels normally into the planar target and the entire process is known as resonance absorption (RA) (22,38). The current pulse is due to RA generated single or multiple collimated relativistic electron jets (39). These jets generate giant, azimuthal magnetic fields ($B_\phi$), having peak pulse height of few hundreds of Megagauss with typical pulse widths of a few ps (19-21) (Schematic in Fig. 3C and details on RA in supplementary)

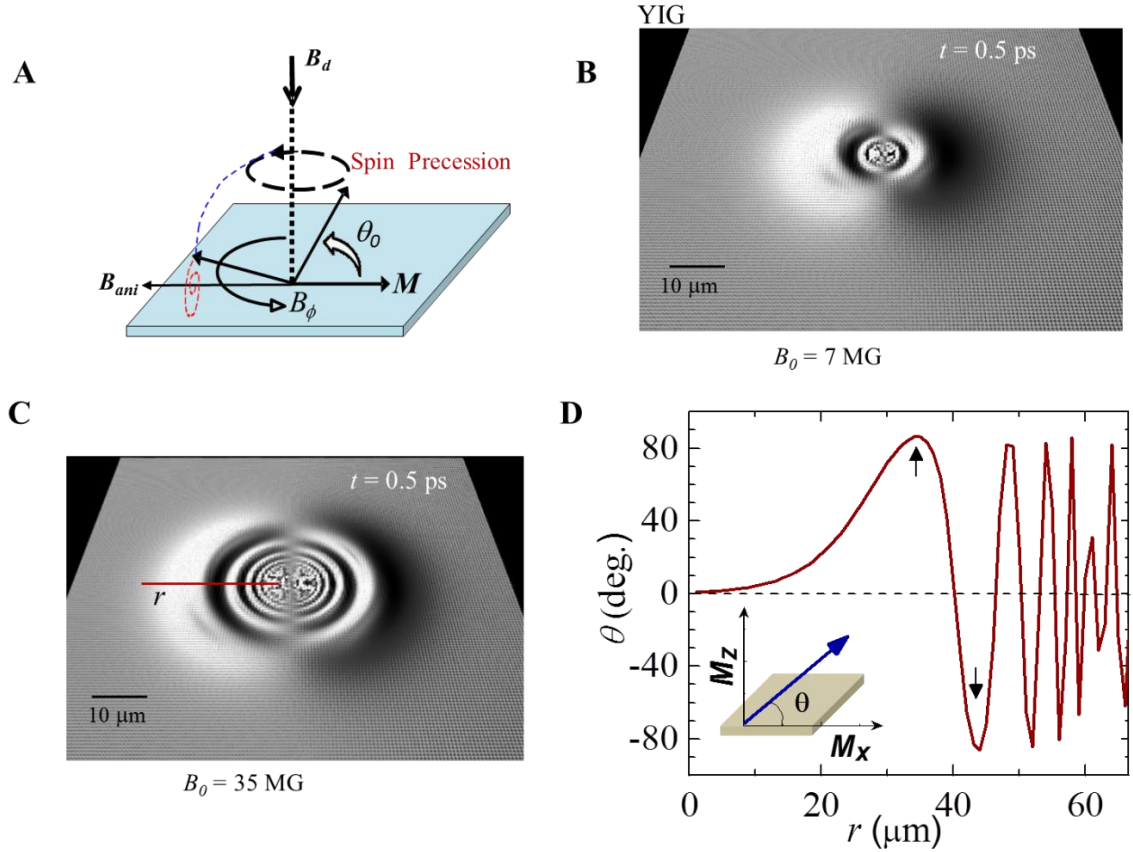

**Fig. 2.** Simulation of YIG films using *MuMax*. (*A*) Schematic of the magnetic precession of the in-plane moments of magnetic film due to the megagauss magnetic field (see text for details). (*B,C*) Simulation of YIG films showing ripples in the magnetic moment configurations are generated in the YIG film with magnetic pulses, $B_0$ = 7 megagauss and 35 megagauss, recorded at time (*t*) = 0.5 secs (see text for details). The ripples spread out as a function of increasing time. (*D*) Plot of $\theta = \tan^{-1}\left(M_z/M_x\right)$ across the red solid line in (*D*) shows periodic modulations of the magnetic moment configuration in the film at *t* = 0.5 ps.

To explore the effect of these RA generated giant magnetic field pulses on YIG films, we model the temporal evolution of in-plane local magnetization $\vec{M}$ in the YIG film under the influence of a magnetic field pulse (refer Fig. 2*A*) through the Landau-Lifshitz-Gilbert (LLG) equation using the MuMax software (40,41) (see Methods for simulation details)

$$\frac{d\vec{M}}{dt} = \gamma\left(\vec{B}_{eff} \times \vec{M}\right) + \frac{\alpha}{M_s}\vec{M} \times (\vec{M} \times \vec{B}_{eff}) \quad\quad\quad\quad\quad eqn.\ 1$$

The first term on the right in *eqn.*1 is the torque on $\vec{M}$ due to the effective magnetic field ($\vec{B}_{eff}$ = Applied in plane azimuthal field ($\vec{B}_\phi$) + Demagnetizing Field ($\vec{B}_d$) + Anisotropy field ($\vec{B}_{ani}$)), where γ is the gyromagnetic ratio. The in-plane azimuthal field pulse (see Fig. 2A) is $\vec{B}_\phi = B_\phi \hat{\phi}$ where $\hat{\phi}$ is the azimuthal unit vector in film plane. The second term is a damping term with damping constant α. The $\vec{B}_\phi$ pulse first causes $\vec{M}$ to flip out of the film plane by an angle $\theta_0$ due to a torque, $\vec{\tau} = \vec{M} \times \vec{B}_\phi$. For YIG a large $B_\phi$ pulse of 7.5 MG (discussed subsequently) gives a $\theta_0 = (\gamma B_\phi)\Delta t = 10.5°$ in a pulse duration $\Delta t \sim 0.5$ ps (YIG has γ = 28 GHzT$^{-1}$ (42)). The flipped $\vec{M}$ generates a demagnetization field ($\vec{B}_d \sim -\vec{M}\sin\theta_0\ \hat{z}$, $\hat{z}$ is ⊥ to film) (see Fig. 2A) around which $\vec{M}$ precesses. At this stage, spin waves are excited in the magnetic material. Simultaneously, damping (second term in *eqn.*1) comes into play, leading to the decay of the waves as the precessing $\vec{M}$ gradually loses energy and falls towards the film plane (see blue dashed trajectory in Fig. 2A), until its energy is just lower than the anisotropy energy barrier (∝ $K_u \sin(\psi)$, ψ - angle between $\vec{M}$ and $\vec{B}_{ani}$). Note that the anisotropy energy is minimum for parallel (ψ = 0) or antiparallel (ψ = π) orientation. Here $\vec{M}$ performs a damped precession around $\vec{B}_{ani}$ before falling back into the film plane, with $\vec{M}$ oriented either along or opposite to $\vec{B}_{ani}$. Finally, minimizing the free energy leads to domains with in-plane $\vec{M}$, where the in-plane $\vec{M}$ orientation periodically flips by $\pi$ across the domains.

There appears to be only one other group that has studied the influence of ultra-strong, ultrashort magnetic pulses on magnetic films (43,44) and it is therefore interesting to make a comparison with their results. Their studies subjected films of high $K_u$ ($K_u \sim 0.1$ MJ m$^{-3}$) viz., high $\vec{B}_{ani}$ to (a) magnetic field pulses of the order of a few tens of Tesla and also (b) electric (*E*) fields ~$10^9$ V/m. These fields were generated by irradiating the film directly with relativistic electron (e⁻) bunches with 28 and 40 GeV energies at SLAC (43,44). The presence of a strong anisotropy field in the magnetic film material resulted in anisotropic dumbbell like domain pattern (43,44). The in-plane $\vec{M}$ across adjacent domains in the dumbbell pattern were anti-parallely aligned. A comparison with our work and shows the distinct spatial symmetry of our domain shapes, viz., layered onion shell structure, which is in contrast to the anisotropic dumbbell shaped domains found in the SLAC studies. Another distinguishing feature of our domains is that - as one moves along a line cutting across the domains the $\vec{M}$ periodically twists out of the plane leading in turn to a periodic modulation of local $B_z$ (see Fig. 1*E*), while in the SLAC study, $\vec{M}$ always remains in-plane. These contrasting results suggest significant differences in the physical processes leading to the distinct domain shapes seen in our experiments. We would like to suggest that our use of YIG has properties which are completely different from those used in the SLAC study. YIG has a time independent damping constant ($\alpha$), which is nearly two orders of magnitude smaller compared to that of the material used in the SLAC study (see supplementary information). Furthermore, the domain structures seen in the SLAC study are only explained by considering that for the materials used in their study, the $\alpha$ increase exponentially with time until, the spin wave carries away all the energy from the precessing $\vec{M}$. For YIG, not only is $\alpha$ low but also we do not need to consider any time dependence of $\alpha$ to explain the domain feature we observe. Due to low $\alpha$ value, for our YIG simulations the damping (second) term in *eqn.*1 has a relatively small effect (compared to that in the SLAC studies of refs. 43,44) resulting in stable spin waves excited by the field pulse which propagate across the YIG film. Also note that for YIG, $K_u$ = 6.1 x $10^{-4}$ MJ m$^{-3}$ is nearly two orders smaller compared to the strong magnetic anisotropy material used at SLAC (43,44), hence the domains formed in YIG are more symmetrical compared to the asymmetrical dumbbell pattern found in the SLAC study. It is also important to mention the differences in the methods used to generate the field pulses in both cases. Our intense laser generated field pulses are essentially magnetic while both $\vec{B}$ and $\vec{E}$ pulses are generated by the relativistic e- bunches in the SLAC experiments (43). The electron beam in the SLAC study traverses the film and causes film damage, while such a deleterious effect is completely avoided in our study. Lastly, our $B_\phi$ pulse is larger by at least an order of magnitude (19,20) (~ 100 T) compared to that in the SLAC study ( ~ 10 T) (43).

For our simulation, the azimuthal magnetic field distribution experienced by the YIG film is approximated using an expression similar to that for the field distribution at positions located away from the relativistic electron bunch (45), viz., $B_\phi(x,y,t) = \frac{B_0}{(r/\sigma)} \xi(t-t_p)$ for $r > \sigma$, where, $r$ is the distance from the projected center of laser irradiation on the YIG film and $\sigma$ is the diameter of region around irradiation center within which there are the RA generated current jet(s). We use $\sigma$ as the diameter of the irradiating laser beam. In the region $r < \sigma$, we use a uniform $B_\phi = B_0$. The temporal behavior of the pulse is $\xi(t-t_p) = 1$ for $0 \leq t \leq t_p = 1$ ps and $\xi(t-t_p) = 0$ for $t > t_p$. In the above expression, $B_0 = \frac{2\mu_0 n e}{(2\pi)^{3/2} t_p \sigma}$, is the field (45) present outside the boundary of radius $\sigma/2$, where $n$ is the number of electrons in the current jet and $e$ is the electron charge. Figures 2*B-C* show results of the LLG simulations in YIG with a time independent damping constant α, at time (*t*) = 0.5 ps after the application of magnetic field pulse with peak field $B_0$ of 7 MG and 35 MG respectively. Note that $B_0 = 7.5$ MG ($n = 1.2 \times 10^{12}$) and 35 MG ($n = 5.6 \times 10^{12}$) correspond to an order of magnitude larger number of electrons in our intense laser pulse generated electron jet compared to those in the relativistic electron bunch at SLAC. The Zeeman energy associated with our giant $B_\phi$ pulse in YIG is ~ 14296 MJ m$^{-3}$ (see supplementary section), is sufficiently large to completely overwhelm the low magnetic anisotropy energy of YIG ($K_u = 6.10 \times 10^2$ J m$^-$

$^3$). Figures 2*B* and *C* show that the $B_\phi$ pulse excites circular spin wave ripples in the YIG film around $\vec{r}_o$ (center of the pulse). With increasing time, the ripples spread out across the film (see movie (46)), until they reach the film edge which occurs within 1 ns. At long times (~ 100 ms), a complex rectangular multi-domain configuration is stabilized in the YIG film, which are unlike the domains we have recorded in Fig. 1 (see movie at link Ref. 46, and supplementary information section). We do not observe any dependence of our simulation results on the lateral film dimensions as long as they are larger than the ripple wavelength. The rippling feature seen in the simulations closely resembles our observed elliptical layered onion shell domain structure of Fig. 1. Further similarity between the simulated features and our experiment is seen in Fig. 2*D*, where we calculate the orientation ($\theta$) of local $\vec{M}$ in the film w.r.t the film plane viz., $\theta = \tan^{-1}\left(M_z/M_x\right)$, measured along the radial direction drawn as a red line in Fig. 2*C*. The contrast modulations seen in Figs. 2*B* and *C* clearly correspond to the periodic flipping of *z* component of $\vec{M}$, which is evident from the $\theta(r)$ behavior in Fig. 2*D*. The periodic flipping of $\vec{M}$, viz., the behavior of $\theta(r)$ in Fig. 2*C* is similar to the behavior of $B_z(r)$ in Fig. 1*E*.

Figures 2*B*-*C* show that at 0.5 ps, the diameter of the outer edges of the rippling structure are comparable for $B_0 = 7.5$ MG and 35 MG. The number of rings inside the concentric circular structures increases with $B_0$. These observations from the simulations match with that of Fig. 1- while the overall size of the domain is independent of the laser intensity the number of concentric rings (*N*) increase with intensity (cf. Figs. 1*B* and 1*F*). The simulations show that increase in *N* is related to increase in the pulse peak field $B_0$ (cf. Figs. 2*B* and 2*C*). By comparing the number of rings (*N*), determined as a function of laser intensity (*I*) (experiments, like Figs. 1*B*, *D* and *F*) and as a function of $B_0$ (determined from simulations like those in Fig. 2), an empirical relation between $B_0$ and *I* is obtained (details in supplementary). Our experiments reveal that for $I \leq 10^{17}$ Wcm$^{-2}$, no elliptical domains form. Using the empirical relation, this intensity corresponds to a peak field of $B_0 \sim 0.1$ MG. Our simulations also confirm that there are no discernible magnetized ripples excited in YIG below 0.1 MG field.

Simulating *eqn.* 1 using a general form of $B_\phi$ with multi-peaks (19-22), viz., $B_\phi(x,y,t) = \sum_i \frac{B_{0,i}}{(r_i/\sigma)} \xi(t - t_{p,i})$ with $i = 1$ to 2, we observe in Fig. 3*B* two well separated rippling structures (after 0.5 ps), which are similar to the features in Fig. 1*F*. By reducing the spacing between the field pulses down to 18 μm we observe (Fig. 3*C*) a single rippling structure produced by the overlap of two spin wave ripples. The resulting ripple is not circular but elliptical in shape. It is these elliptical shapes that we observe in our experiments (Fig. 1). The multi-peaked field structure may result from multiple closely spaced e-jets generated during RA process as shown in the schematic of Fig. 3*A*.

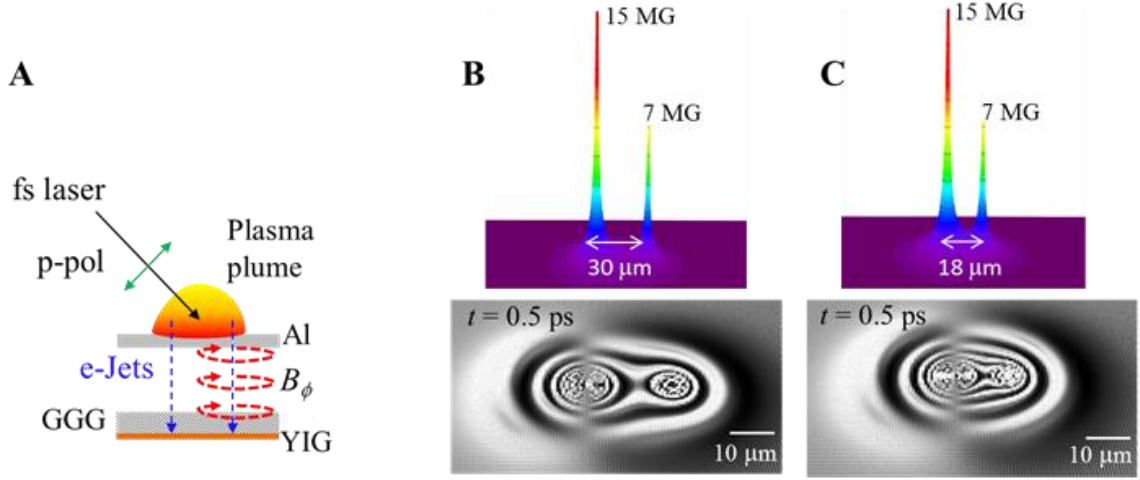

**Fig. 3**. Correlation of simulations with experiments. (*A*) Schematic of the Resonance Absorption processes in which fs laser hits the Al film to form a plasma plume and interactions with the laser plasma creates multiple electron (e) jets. These e-jets moving at relativistic speeds carrying mega-amps of current generate magnetic field around it ($B_\phi$) which is shown only around one of the e-jet paths for the sake of clarity. These electron jets give rise to megagauss azimuthal magnetic fields inside the YIG. *(B,C)* Simulations done on the YIG film using two sharply peaked magnetic field pulses having $B_0$ of 15 MG and 7 MG, separated by 30 μm and 18 μm. The ripples created in the YIG film are well separated when the separation between the field pulses is significant while the two ripples merge into an elliptical shaped ripple when the pulses are closer to each other. The ripple patterns shown are obtained by stopping the simulation at $t$ = 0.5 ps.

From our simulations, we determine the phase velocity ($v_p$) of the spin waves generated by the giant field pulse, as the ratio of the distance covered by the crest of a ripple to the time taken. The $v_p$ turns out to be in the range of ~ $10^7$ m s$^{-1}$ which is four orders of magnitude greater than the typical limit of domain wall velocity (~ 1000 ms$^{-1}$) reported for YIG films (47). Note that $v_p$ is not related to physical motion of domain walls. Using $v_p$ and the measured wavelength of the ripple ($\lambda$) excited at different energies, we obtain $\omega_m(k) = v_p \frac{2\pi}{\lambda}$ for our spin wave to be in the range of ~ 30 to 100 THz (depending on the laser intensity). Some earlier measurements suggested that spin waves with frequencies $\omega_m$ ~ few THz (48,49) can be excited in YIG films. We show that one to two order of magnitude higher THz frequency spin waves can be excited using our intense laser and the novel metal/dielectric/YIG film sandwich target combination.

The close match between our simulations and experiments in YIG demonstrates a unique effect, viz., excitation of fast propagating spin waves excited in the YIG film. Furthermore, the YIG effectively captures a snapshot of the propagating spin wave. It is pertinent to ask by what mechanism is the spin wave transforming into a static magnetic domain pattern? To get a glimpse into the answer, we determine the amount of non-collinear $\vec{M}$ configuration generated at the domain site $i$ in YIG by the giant field pulse viz., the average value of $|\vec{M}_i \times \vec{M}_j|$ within a region of 10 μm × 10 μm inside the simulated ripple of magnetic disturbance in Fig. 2. The average value of $|\vec{M}_i \times \vec{M}_j|$ per unit area associated with each ripple turns out to be ~ 0.9 μm$^{-2}$ compared to zero μm$^{-2}$ prior to the pulse. We propose that the spin waves excited in YIG generate a non collinear twisted moment configuration which leads to an enhanced contribution of magnetic energy associated with the spin configuration in YIG coming from the Dzyloshinskii-Moriya (DM) (5,6) interaction energy, where DM $\propto |\vec{M}_i \times \vec{M}_j|$. The additional DM interaction in YIG excited by the spin wave we believe stabilizes the non-collinear magnetization configuration resulting in the spin wave ripple pattern being frozen into YIG as the elliptical, concentric onion shell-like domains as seen in Fig. 1. It may be recalled that DM interactions in conventional condensed matter systems are crucial for stabilizing the small chiral magnetic structures like skrymions (4). The ever present pinning in the YIG film would also contribute to stabilizing the *M* configuration excited by the spin wave. Pinning of the outermost spin waves ripples by microscopic defects in the film is responsible for the irregular shape of the outermost ring seen in the images of Fig. 1.

From the layered onion shell domains recorded in YIG film we show that femtosecond laser pulse excites THz spin waves in YIG with diffusion lengths upto hundreds of microns. This result has potential applications for spin wave based devices. In recent times, the generation, control and manipulation of spin waves have emerged as an important research area (50,51) as they can be utilized in dissipation-less transfer of information across a device. In conventional electron transport based devices, increasing electrical resistance with miniaturization significantly increases the dissipation losses, thereby limiting the processing speed of these devices. However in magnonic devices, spin waves with frequencies, $\omega_m <$ 10 GHz, are launched via spin Hall effect or spin torque effect using an electrical current (27,34). The electrical contacts in these magnonic devices for current injection result in unavoidable Joule heating dissipation at the contacts. These devices also need to be operated at low temperatures as the spin waves are susceptible to thermal noise, since for $\omega_m <$ 10 GHz the $\hbar\omega_m << k_BT$ (=0.025 eV at 300 K). However, the spin waves excited in our metal/dielectric/YIG sandwich using indirect irradiation of YIG with the intense laser pulses, completely eliminate the need for using electric currents for exciting the waves and hence limit Joule heating dissipation. Furthermore, the fast propagating spin waves with long diffusion lengths reaching up to hundreds of microns have frequencies that can be varied between 10 to 100 THz by varying the laser intensity. Hence magnonic devices employing our method are suitable for room temperature operation, as the energy of spin waves excited in YIG by intense laser pulse is far above the thermal noise floor (as for $\omega_m$ ~10 to 100

THz, $\hbar\omega_m \geq k_B T$ ). We also re-emphasize that our design for launching spin waves has multiple advantages compared to doing the same with an accelerator based relativistic electron beam (43,44). We have a much more compact table top design, less complexity in operation and the strength of the field pulses and pulse durations can be conveniently controlled by varying the laser intensity, the target design and the interaction geometries.

**Conclusions**

We have demonstrated novel, macroscopic, concentric elliptical onion shell-like magnetic domains created in Yttrium Iron Garnet (YIG) film via giant magnetic field pulses of picosecond duration, generated from femtosecond, intense laser pulses. Spin waves excited in the YIG film by the field pulse generate non-collinear spin configurations in the YIG film which gives rise to DM interactions which initially was very weak in the YIG film. These interactions together with pinning in these films cause the spin waves to stabilize into surprisingly long lived elliptical domain patterns. Our innovative metal/dielectric/YIG structure irradiated with intense laser pulse offers a new way to excite ultrafast spin waves with frequency in the few tens to hundreds THz range with large diffusion lengths. These high frequency spin waves are undisturbed by thermal fluctuations. Hence our tabletop route offers a new way for developing a dissipation less high frequency magnonic devices which are capable of operating at room temperature (52).

**Acknowledgements:** SSB would like to acknowledge the funding support from IIT Kanpur and the Department of Science and Technology, Government of India, New Delhi. GRK acknowledges partial support from the Science and Engineering Board (SERB), Government of India, New Delhi through a J C Bose National Fellowship (JCB-2010/037). GRK acknowledges late Predhiman K. Kaw for stimulating discussions.

**Methods:**

**Simulations using MuMax:** We simulate YIG films with dimensions $150 \times 150$ $\mu m^2$ and thickness 3 $\mu m$. The material parameters used in the calculations are as follows: $M_s = 1.4 \times 10^5$ A/m, exchange constant $A = 3.6 \times 10^{-12}$ J/m, α = 0.0005 and anisotropy constant $K = 610$ J/m$^3$. The film is discretized cells with grid lengths of 1 $\mu m \times 1$ $\mu m \times 1$ $\mu m$. At the vertex of each cell a magnetic moment is placed. It may be noted that due to the computationally intensive nature of the simulations, we simulate a structure smaller than that in the experiment. We have however checked that the main features of the simulations remain the same irrespective of the film size used for simulations. Also, we were interested in capturing the long wavelength magnetic modes which maybe excited on the YIG films by the application of giant magnetic field pulses, rather than the excitation of small (submicron) wavelength modes. The initial magnetic configuration of YIG in the simulations is in-plane. We have used magnetic field pulse of pulse height varying between 0.07 Mega-Gauss (half-maxima) and 350 Mega Gauss, pulse width σ = 10 μm and pulse time, $t = $ 1ps (see text for details).

Figure 1

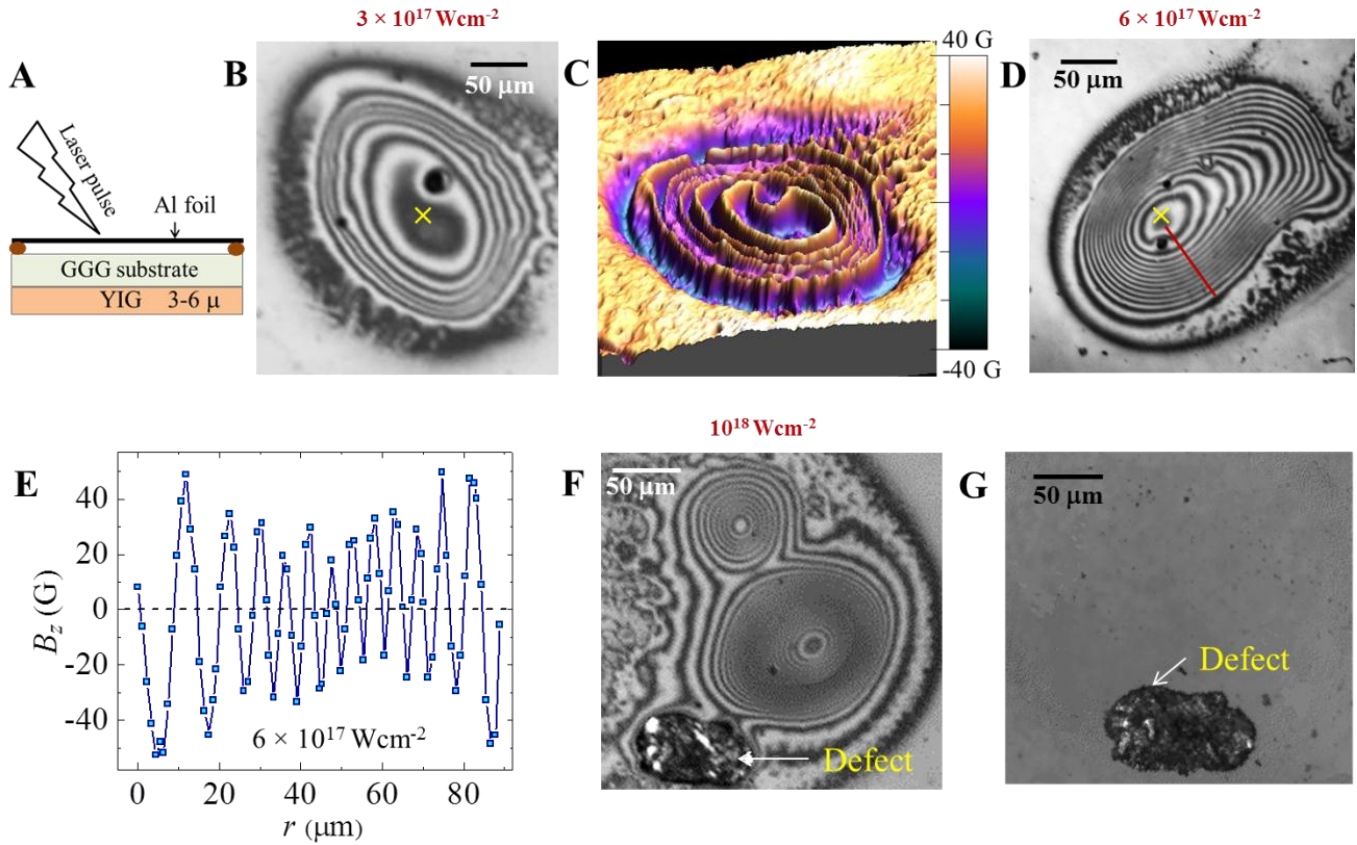

**Fig. 1,** MOM Images of laser Irradiated YIG films. (*A*), Schematic shows the incidence of the femtosecond laser pulse on a 16 μm thick Al film suspended 100 μm above the GGG substrate, on the lower side of which is the YIG film. A dielectric layer of thickness ~ 200 μm composed of air and the GGG substrate, exists between the Al and the YIG film. The × mark is the estimated projected location of the laser spot on the YIG film. (*B,D,F*) MOM images of concentric elliptical magnetic domains generated in YIG film are shown after irradiation at laser intensities (I) of $3 \times 10^{17}$ Wcm$^{-2}$, $6 \times 10^{17}$ Wcm$^{-2}$ and $10^{18}$ Wcm$^{-2}$, respectively. (*C*) Color coded three-dimensional map of the $B_z(x,y)$ distribution in the elliptical domains of 1(*B*). The colors represent the magnitude of the $B_z$ values as shown in the color-bar scale. (*E*) A one-dimensional map of $B_z$ profile (viz., $B_z(r)$) measured along the red line in Fig. 1(*D*), shows periodic oscillations in $B_z$ as one traverses the concentric rings of the elliptical domain pattern. The maximum amplitude of the oscillating $B_z(r)$ is ± 40 G. (*F*) Shows two domain patterns generated in YIG with laser pulse irradiation of intensity of $10^{18}$ Wcm$^{-2}$. The smaller upper domain structure is more circular than the one below. Also seen is a defect in the YIG film which was present before irradiation. (*G*) MOM image of the region same as that in (*F*), imaged after 10 days of laser irradiation, with the black defect as the identifier. Here we see layered onion shell like magnetic domain patterns have disappeared.

Figure 2

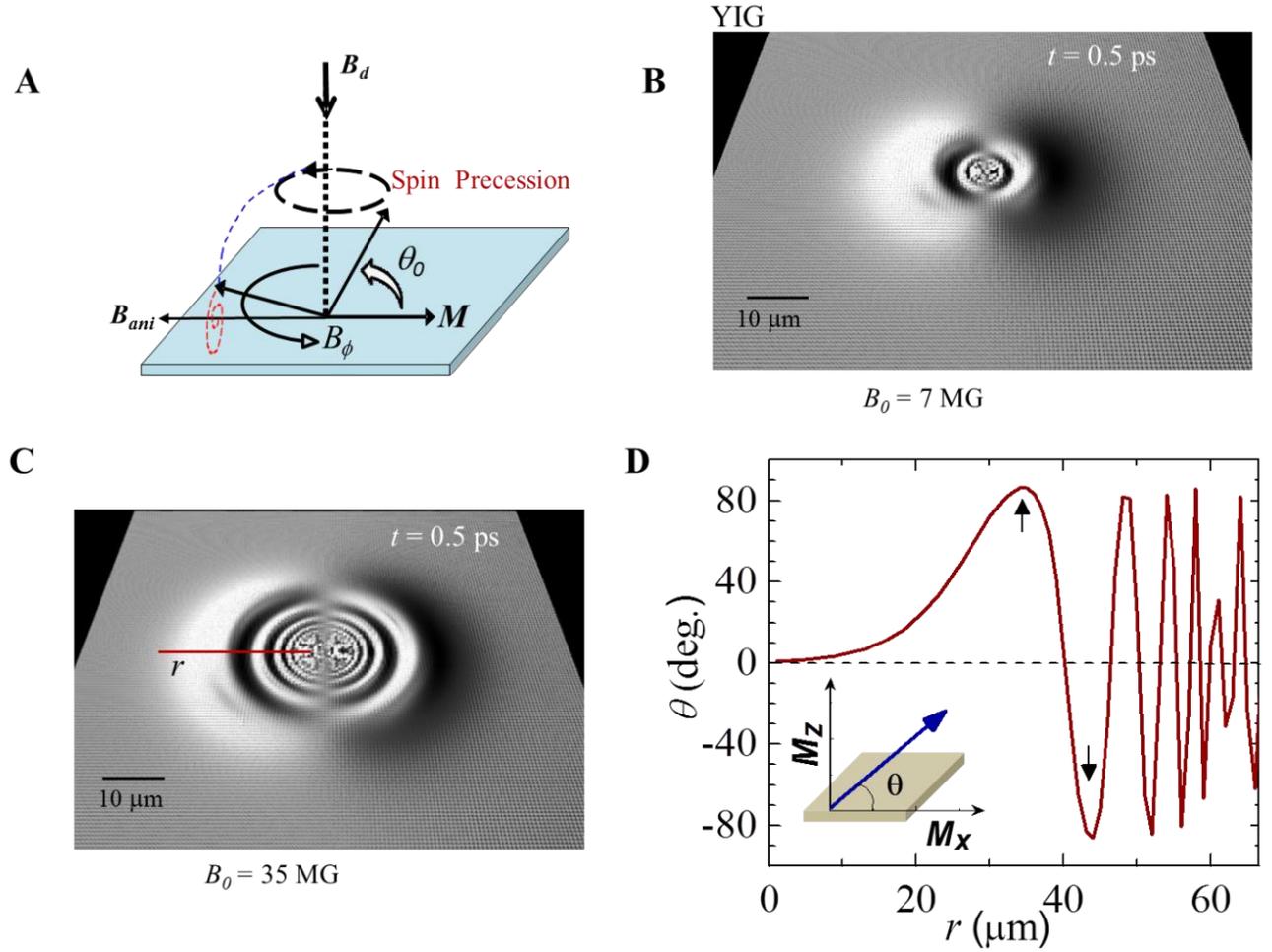

**Fig. 2.** Simulation of YIG films using *MuMax*. (*A*) Schematic of the magnetic precession of the in-plane moments of magnetic film due to the megagauss magnetic field (see text for details). (*B,C*) Simulation of YIG films showing ripples in the magnetic moment configurations are generated in the YIG film with magnetic pulses, $B_0$ = 7 megagauss and 35 megagauss, recorded at time (*t*) = 0.5 secs (see text for details). The ripples spread out as a function of increasing time. (*D*) Plot of $\theta = \tan^{-1}\left(M_z/M_x\right)$ across the red solid line in (*D*) shows periodic modulations of the magnetic moment configuration in the film at *t* = 0.5 ps.

Figure 3

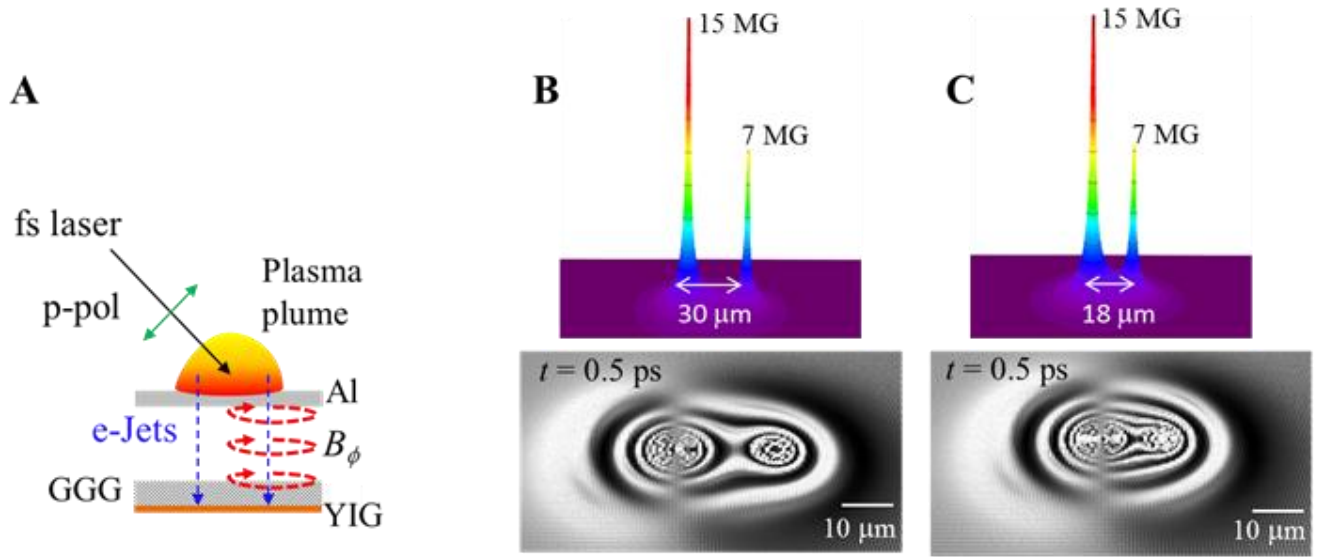

**Fig. 3**. Correlation of simulations with experiments. (*A*) Schematic of the Resonance Absorption processes in which fs laser hits the Al film to form a plasma plume and interactions with the laser plasma creates multiple electron (e) jets. These e-jets moving at relativistic speeds carrying mega-amps of current generate magnetic field around it ($B_\phi$) which is shown only around one of the e-jet paths for the sake of clarity. These electron jets give rise to megagauss azimuthal magnetic fields inside the YIG. *(B,C)* Simulations done on the YIG film using two sharply peaked magnetic field pulses having $B_0$ of 15 MG and 7 MG, separated by 30 μm and 18 μm. The ripples created in the YIG film are well separated when the separation between the field pulses is significant while the two ripples merge into an elliptical shaped ripple when the pulses are closer to each other. The ripple patterns shown are obtained by stopping the simulation at *t* = 0.5 ps.

# Supplementary Materials for

# Macroscopic, layered onion shell like magnetic domain structure generated in YIG film using ultrashort, megagauss magnetic pulses

**Kamalika Nath[1], P. C. Mahato[1], Moniruzzaman Shaikh[2], Kamalesh Jana[2], Amit D. Lad[2], Deep Sarkar[2], Rajdeep Sensarma[3], G. Ravindra Kumar[2,#], S. S. Banerjee[1,*]**

[1]Department of Physics, Indian Institute of Technology, Kanpur, Uttar Pradesh, India

[2]Department of Nuclear and Atomic Physics, Tata Institute of Fundamental Research, Homi Bhabha Road, Mumbai, India

[3]Department of Theoretical Physics, Tata Institute of Fundamental Research, Homi Bhabha Road, Mumbai, India

Corresponding authors' Email: [#]grk@tifr.res.in, [*]satyajit@iitk.ac.in

## Materials and Methods

### Simulations using MuMax

We simulate YIG films with dimensions $150 \times 150$ $\mu m^2$ and thickness 3 $\mu m$. The material parameters used in the calculations are as follows: $M_s = 1.4 \times 10^5$ A/m, exchange constant $A = 3.6 \times 10^{-12}$ J/m, $\alpha = 0.0005$ and anisotropy constant $K = 610$ J/m$^3$. The film is discretized cells with grid lengths of 1 $\mu m \times 1$ $\mu m \times 1$ $\mu m$. At the vertex of each cell a magnetic moment is placed. We use large grid lengths in our simulation as we are interested to capturing long wavelength magnetic modes which maybe excited on the YIG films by the application of giant magnetic field pulses. We are not interested in the excitation of small (submicron) wavelength modes. The initial magnetic configuration of YIG in the simulations is in-plane. We have used magnetic field pulse of pulse height varying between 0.07 MG (half-maxima) and 350 MG , pulse width $\sigma = 10$ $\mu m$ and pulse time, $t = 1$ $ps$ (see text for details).

**Experimental set-up**

The laser irradiation experiments are performed in experimental chamber with base vacuum of $10^{-5}$ mbar. The experimental schematic is shown in Fig. S1 (a). To irradiate a fresh portion of sample every laser shot, the sample is mounted on precise X-Y-Z-θ stage assembly. *P*-polarized 25 fs, 800 nm laser pulses were focused to 20 μm diameter spot by off-axis parabolic mirror (OAP) on to a sample to create a plasma. The intensity on the sample is varied from $10^{17}$ to $10^{18}$ Wcm$^{-2}$, by changing the laser energy appropriately. The angle of incidence on sample is maintained to 45° to maximize resonance absorption (RA) (discussed later).

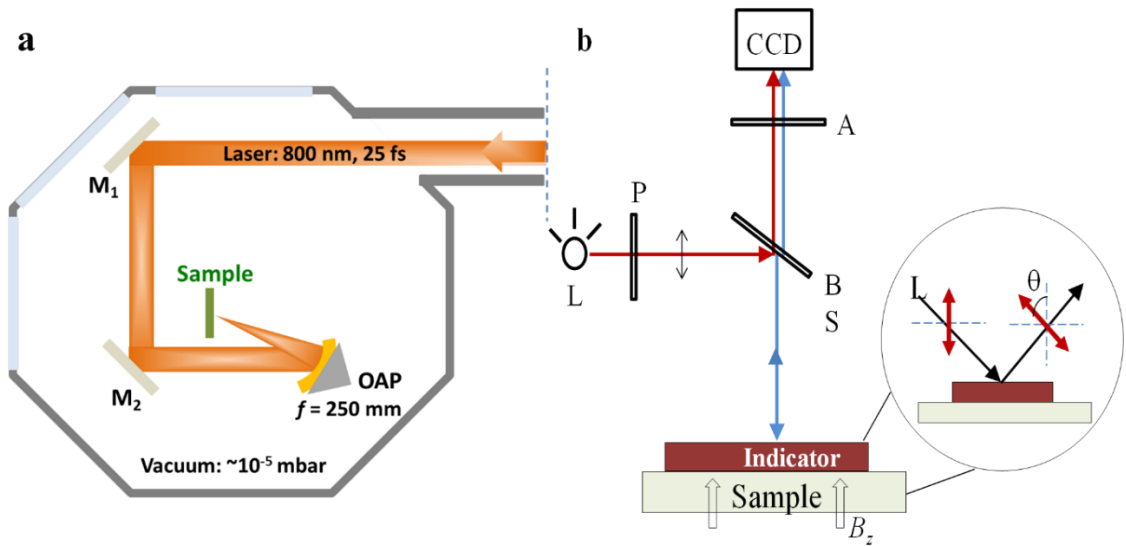

**Figure S1: (a)** Schematic of intense, femtosecond laser irradiation set-up. M1, M2: Mirrors, OAP: off-axis parabolic mirror. **(b)** Schematic of Magneto-optical Imaging set-up.

The laser irradiated samples are imaged by high sensitivity magneto-optical imaging technique, which is based on the principle of Faraday effect. Figure S1(b) shows a schematic of the magneto-optical setup where the components refer to: un-polarized Light source (L), Polarizer (P), Beam Splitter (BS), Analyzer (A). In Faraday rotation, the plane of polarization of a linearly polarized light undergoes rotation by an angle $\theta$ in presence of magnetic field in the direction of light propagation. The Faraday rotated angle ($\theta$) is given by, $\theta = V\mathbf{B}d$, where $V$ is the Verdet constant (depends on material and the wavelength of light), $B$ is the magnetic field and $d$ is the path length in the sample. Therefore, we get information about the magnitude and direction of the local magnetic field from the sample by knowing the degree of rotation $\theta$. The Faraday rotated intensity distribution is captured in a CCD camera and henceforth calibrated to the corresponding $B$ values. In a MOI, the intensity of the Faraday rotated light $I(x,y)$ is proportional to the local field $B_z^2$. We calibrate the $I(x,y)$ versus well calibrated magnetic field by applying known fields to the YIG film. Using this information, we convert the $I(x,y)$ in the MO images of Fig. 1 (main text) into $B_z(x,y)$. The bright and dark contrasts in the MO images represent the local $B_z(x,y)$ fields and hence $M_z$ pointing either out or into the YIG film respectively. The shade in the image corresponds to the magnitude of $B_z$.

**Resonance absorption**

We give a short summary of the generation and effects of magnetic fields as a result of resonance absorption (RA) in laser-matter interaction. During the RA process, the incoming *p*-polarized (*E*-field in the plane of incidence) high intensity laser pulse impinges on the sample at oblique incidence and generates dense plasma. The preformed plasma created by the laser pre-pulse expands away from the target surface as shown in the schematic (Fig. S2). The laser light propagating into such plasma encounters a density gradient in the plasma, with the density being highest close to the site of irradiation. The electrons of mass $m_e$ and charge $e$ in the plasma with a density $n_e$ oscillates with a frequency $\omega_p$, known as the plasma frequency. If the incident light field has a frequency $\omega < \omega_p$ then the light is reflected from the plasma. Inside the plasma dome the density increases as one approach the surface of the material. Hence, the femtosecond laser beam is being reflected from the 'critical surface' (where the local plasma frequency matches that of the light wave). However, an evanescent wave continues deeper into the plasma.

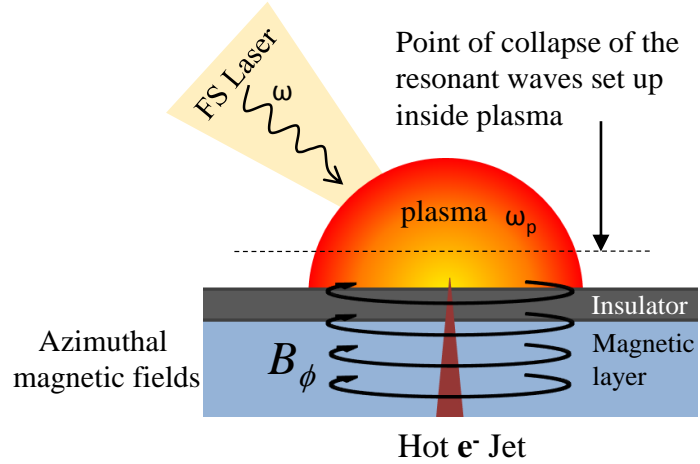

**Figure S2:** Schematic of Resonance Absorption mechanism during fs laser-matter interaction.

up to a region inside the plasma with a critical density. If the frequency of the evanescent light field (ω) matches the plasma frequency ($\omega_p$) in the dense region of the plasma then large resonant amplitude oscillations are excited in the plasma. Due to the collapse of the resonant waves in the critical layer of the plasma, energy is transferred from the plasma wave energy to the electrons in the plasma, thereby generating a collimated jet of hot electrons channeling through the target material at relativistic speeds. Consequently giant, megagauss range azimuthal magnetic field (*viz.*, fields in the plane of the target material and perpendicular to the electron jet) are generated from the strong mega ampere currents associated with the hot electrons jet. While these magnetic fields ($B_\phi$) are very large, their typical lifetime is a few picoseconds. Thus resonance absorption leads to generation of megagauss magnetic field pulses for picosecond time interval.

**Relation between *P* and $B_0$:**

The log-log plots in Figs. S3(I), (II), show a linear relation between *N* and *P* and *N* with $B_0$. The trends of the experimental data and simulation results of *N* vs *P* and $B_0$ behave linearly in the double log plot suggesting $N = C_1 P^{\alpha}$ and $N = C_2 B_0^{\beta}$. From the same *N* using the values of *P* and $B_0$ deduced from panels I and II we plot on a double log scale the linear relationship between $B_0$ and *P* which is displayed on a log – log plot in Fig. S3(III). From the fit in the panel we obtain an empirical relation $B_0 \propto P^{3.5 \pm 0.1}$ and this empirical relation is approximately valid until the ripples reach the edges of the film. Using this empirical relationship, we estimate that our laser intensity $P \sim 7 \times 10^{17}$ W cm$^{-2}$ corresponds to generating a gigantic field pulse $B_0 \sim 100$ MG experienced by the YIG. We would like to mention that we did not observe formation of any of the elliptical domain structures below a lower threshold of $P = 10^{17}$ W cm$^{-2}$, which from backward extrapolation in Fig. S3(III) corresponds to $B_0 \sim 0.1$ MG. This estimate concurs well with our simulations which show no excitation of long wavelength ripples for $B_0 < 0.07$ MG.

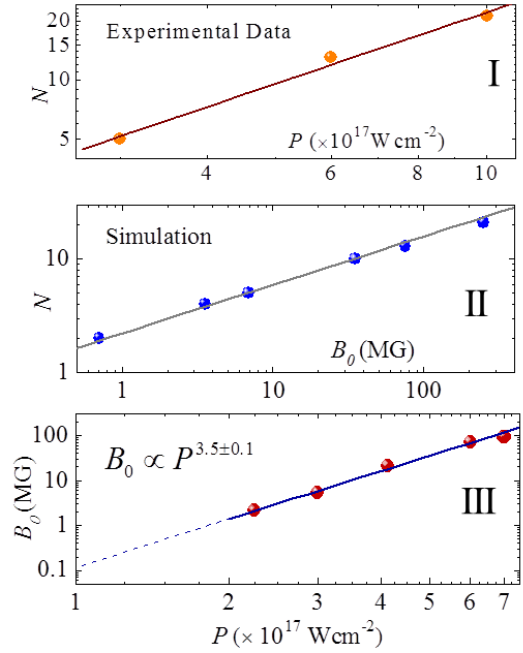

**Figure S3:** Panel I shows a plot of *N vs P* from experiments and panel II shows a plot of *N* vs $B_0$ obtained from simulations (both on log-log scale). Panel III shows the correlation between *P* and $B_0$.

**A comparison of material parameters between YIG and the material studied in Ref. [i] (Ref. [43] in main MS), viz., $Co_{70}Fe_{30}$ film.**

| Material Parameters | YIG | $Co_{70}Fe_{30}$ |
|---|---|---|
| Magneto-crystalline Anisotropy ($K_u$) | $6.10 \times 10^2$ J.m$^{-3}$ | $7.6 \times 10^4$ J.m$^{-3}$ |
| Damping Constant ($\alpha$) | 0.0005 | 0.015 |

A comparison from the above table shows that the $K_u$ and $\alpha$ of the $Co_{70}Fe_{30}$ films used in the SLAC study [i] is two orders of magnitude larger than those in YIG.

YIG has a cubic lattice with lattice parameter of $a$ = 12.37 A°, magnetic moment of $\mu$ = 40 $\mu_B$ per unit cell [ii]. The Zeeman energy density of YIG associated with giant magnetic field pulses, $B_\phi \sim$ 7.5 MG can be written as, $\frac{\mu B_\phi}{a^3} = 14296$ MJ.m$^{-3}$. This Zeeman energy ($E_z$), generated either by the relativistic electron bunch at SLAC [43] or via intense FS laser pulses, overwhelm the magnetic anisotropy energy barrier of materials (for YIG $E_z \sim$ 14296 MJ.m$^{-3}$ >> its magnetic anisotropy energy, $K_u$ = 6.10 x 10$^2$ J.m$^{-3}$).

**Shape of domains formed in YIG after a long time (~ 100 ms after field pulse)**

In YIG, the equilibrium (100 msecs after $B_\phi$ pulse) domain configuration is rectangular shaped, which minimizes the free energy (Fig. S4). These are the natural equilibrium shapes of the domains in YIG which minimize the free energy of the domains. Note the layered onion shell like magnetic domains in YIG film (Fig. 1 of our main MS) are completely different from these equilibrium rectangular shaped domains. The layered onion shell like magnetic domain structure is a snapshot of a rippling spin wave excited in YIG film by the intense fs laser pulse.

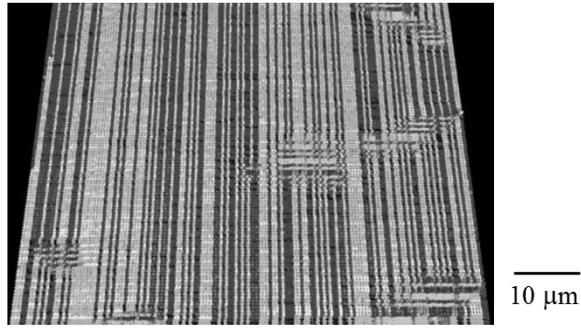

10 μm

**Figure S4:** Showing the simulated magnetization configuration of YIG, recorded 100 msecs after the magnetic field pulse.

[i] S. J. Gamble *et al.*, Electric Field Induced Magnetic Anisotropy in a Ferromagnet. Phys. Rev. Lett. **102**, 217201 (2009).
[ii] Sahalos, John N, Kyriacou, George A., Tunable Materials with Applications in Antennas and Microwaves, (Morgan & Claypool Publishers), 2019.